\documentclass[conference]{IEEEtran}
\usepackage[T1]{fontenc}
\usepackage[latin9]{inputenc}
\usepackage{color}
\usepackage{times}
\usepackage{float}
\usepackage{amsmath}
\usepackage{amssymb}
 \makeatletter

\providecommand{\tabularnewline}{\\}
\floatstyle{ruled}
\newfloat{algorithm}{tbp}{loa}
\providecommand{\algorithmname}{Algorithm}
\floatname{algorithm}{\protect\algorithmname}

\usepackage{multirow,algorithmic}

\usepackage{bbm}\usepackage{latexsym}\usepackage{mathrsfs}\usepackage{amsfonts}\usepackage{amsbsy}\usepackage{times}\usepackage{epsfig}\@ifundefined{definecolor}
 {\usepackage{color}}
\usepackage[dvips]{pstcol}\usepackage{times}\usepackage{indentfirst}\usepackage{wasysym}\@ifundefined{definecolor}
 {\usepackage{color}}{}
\usepackage{bm}

\makeatother

\begin{document}

\title{Locally-Optimized Reweighted Belief Propagation for Decoding
LDPC Codes with Finite-Length \vspace{1em}}

\author{\IEEEauthorblockN{Jingjing Liu} \IEEEauthorblockA{Department
of Electronics\\
 The University of York\\
 Heslington, York, YO10 5DD, UK \\
 Email: jl622@ohm.york.ac.uk} \and \and \IEEEauthorblockN{Rodrigo
C. de Lamare} \IEEEauthorblockA{Department of Electronics\\
 The University of York\\
 Heslington, York, YO10 5DD, UK \\
 Email: rcdl500@ohm.york.ac.uk} \and \IEEEauthorblockN{Henk Wymeersch}
\IEEEauthorblockA{Department of Signals and Systems\\
 Chalmers University of Technology\\
 41296 Gothenburg, Sweden \\
 Email: henkw@chalmers.se}}
\maketitle
\begin{abstract}
In practice, LDPC codes are decoded using message passing methods.
These methods offer good performance but tend to converge slowly
and sometimes fail to converge and to decode the desired
codewords correctly. Recently, tree-reweighted message passing methods have
been modified to improve the convergence speed at little or no
additional complexity cost. This paper extends this line of work and
proposes a new class of locally optimized reweighting strategies,
which are suitable for both regular and irregular LDPC codes. The
proposed decoding algorithm first splits the factor graph into
subgraphs and subsequently performs a local optimization of
reweighting parameters. Simulations show that the proposed decoding
algorithm significantly outperforms the standard message passing and
existing reweighting techniques.\\

\end{abstract}


\section{Introduction}

Low density parity check (LDPC) codes are among the most important
capacity-approaching error-correcting codes \cite{Gallager62:IRE}.
They have a long history, deep theoretical underpinning, and have
found practical application in multiple standards \cite{RyanLin}.
The main power of LDPC codes lies in their pseudo-random nature,
assuring good properties for code, and the availability of
low-complexity decoding algorithms. Recently, a great deal of
research has been devoted to the design of LDPC codes with short to
moderate block lengths, which correspond to most of the application
scenarios of these codes in wireless standards \cite{RyanLin}.

Decoding is commonly based on iterative message-passing methods, allowing
local parallel computations. While message-passing decoding leads
to good performance in terms of bit error rate (BER), it suffers from
a number of drawbacks: (i) convergence to a codeword can take many
iterations, especially with low signal to noise ratios (SNR); (ii) convergence
to a codeword is not guaranteed; (iii) LDPC code design is guided
by the decoding algorithm, constraining codes to have large girths.
In this context, the occurrence of short cycles and stopping sets
makes a significant impact on the performance of LDPC codes, and requires
the development of novel decoding strategies that mitigate these drawbacks.

Different approaches have been considered to deal with these issues.
The most prominent approach is linear programming (LP) based
decoding, which, through a relaxation, formulates the decoding
problem as an LP, and has a maximum likelihood certificate property
\cite{feldman2005using}. LP decoders suffer from high complexity
(exponential in the check node degree), unless further relaxations
are employed \cite{BarLiuDraRec:12}. Another line of investigation,
which aims to improve performance while still maintaining the
message passing nature of the decoder, is that of tree-reweighted
message passing decoding. Based on tree-reweighted belief
propagation \cite{Wainright2}, decoding reverts to a tractable
convex optimization problem, iteratively computing beliefs and
factor appearance probabilities (FAPs). These concepts were applied
to LDPC decoding in \cite{Wymeersch2,LiuLam:12} where the FAPs were
optimized in an offline procedure, subject to additional
constraints: in \cite{Wymeersch2}, the FAPs were constrained to be
constant, while in \cite{LiuLam:12}, the FAPs were constrained to
take on two possible values. In both cases, gains with respect to
standard message passing decoding were observed.

In this paper, we continue this latter line of research, and explicitly
optimize the FAPs off-line, without the constraints from \cite{Wymeersch2,LiuLam:12}.
This allows more freedom in the decoding algorithm without additional
online computational complexity. We propose a locally-optimized reweighting
belief propagation (LOW-BP) decoding algorithm that first splits the
factor graph corresponding to the code into subgraphs and then performs local optimization of the reweighting parameters. The proposed LOW-BP
algorithm can mitigate the effects of short cycles and stopping sets
in factor graphs by applying a reweighting strategy per subgraph.
The LOW-BP algorithm is evaluated for regular and irregular LDPC codes.

This paper is structured as follows. Section II briefly describes
the LDPC system model. Section III reviews reweighting strategies
and existing algorithms. Section IV is dedicated to a detailed
description of the proposed LOW-BP algorithm, whereas Section V
presents and discusses the simulation results. Section VI draws
conclusions from the work.

\section{LDPC System Model}

We consider a rate $K/N$ code, with parity check matrix
$\mathbf{H}$, and a corresponding set of codewords $\mathcal{C}$. Note
that $\boldsymbol{x}\in\mathcal{C}$ if and only if
$\boldsymbol{H}\boldsymbol{x}=\mathbf{0}$. Assuming binary
phase-shift keying and transmission over an
additive white Gaussian noise (AWGN) channel, the
received data are described by
\begin{equation}
\boldsymbol{y}=2\boldsymbol{x}-\mathbf{1}+\boldsymbol{n},
\end{equation}
where $\boldsymbol{n}$ is a sequence of $N$ independently
identically distributed (i.i.d.) AWGN samples with variance
$\sigma^{2}$, and $\boldsymbol{x}\in\mathcal{C}$ is the
transmitted codeword. Given $\boldsymbol{y}$, the aim of the
iterative decoder is to recover $\boldsymbol{x}$ in an iterative
fashion until either $\boldsymbol{H}\boldsymbol{x}=\mathbf{0}$ or
the maximum number of decoding iterations is reached. {Iterative
decoding can be interpreted as message passing on a suitable factor
graph, and is often implemented using belief propagation (BP), or a
variation thereof.}

The factor graph corresponding to our model, $\mathcal{G}(V,E)$,
include the check and variable nodes $V=V_{c}\cup V_{s}$, as well as
a set of edges, $E\subseteq V_{c}\times V_{s}$, such that an edge
connecting the check node $c_{i}$ and the variable node $s_{j}$
exists in the factor graph only if the entry $h_{ij}$ of the
parity-check matrix $\boldsymbol{H}$ equals $1$. The decoding
process can be interpreted as finding
$\boldsymbol{\hat{x}}=\arg\max_{\boldsymbol{x}}p(\boldsymbol{x}|\boldsymbol{y})$.
Using Bayes' rule, this a posteriori distribution becomes
\begin{equation}
\centering{p(\boldsymbol{x}|\boldsymbol{y})=\frac{p(\boldsymbol{y}|\boldsymbol{x})p(\boldsymbol{x})}{p(\boldsymbol{y})}},\label{1}
\end{equation}
which then factorizes as
\begin{equation}
\centering{p(\boldsymbol{x}|\boldsymbol{y})\propto{\prod_{n=1}^{N}\phi_{n}(x_{n})\prod_{m=1}^{M}\psi_{m}(x_{C_{m}})}},\label{2}
\end{equation}
where the \textit{potential} function $\phi_{n}(x_{n})$ corresponds
to $p(y_{n}|x_{n})$, and the \textit{compatibility} function
$\psi_{m}(x_{C_{m}})$ corresponds to an indicator
$\mathbbm{1}\{\sum_{n\in C_{m}}x_{n}=0\}$ within a clique
$C_{m}(m=1,2,\ldots,M)$, corresponding to the $m$-th check
\cite{Wainright2,Wymeerschbook}.

\section{Algorithmic Reweighting Strategies and Variations}

When a factor graph contains short cycles, the standard BP algorithm
normally requires a larger number of iterations but may still fail
to converge. To tackle the issue of non-convergence, Wainwright
\textit{et al.} \cite{Wainright2} presented the tree-reweighted
(TRW)-BP algorithm for cyclic graphs, which aims to impose tighter
upper bounds on the log-partition function. Alongside the TRW-BP
algorithm, a multi-objective function for optimizing the FAPs
$\boldsymbol{\rho}=[\rho_{1},\rho_{2},\ldots,\rho_{M}]$, where
$M=N-K$ is also given
\begin{equation}
\begin{split} \mathcal{F}(\boldsymbol{b},\boldsymbol{\rho}) & =\sum_{n=1}^{N}\mathcal{H}(b_{n})-\sum_{m=1}^{M}\rho_{m}\mathcal{I}_{C_{m}}(b_{C_{m}})\\
 & +\sum_{n=1}^{N}\sum_{x_{n}}b_{n}(x_{n})\log\phi_{n}(x_{n})\\
 & +\sum_{m=1}^{M}\sum_{x_{C_{m}}}b_{C_{m}}(x_{C_{m}})\log\psi_{m}(x_{C_{m}}),
\end{split}
\label{objective}
\end{equation}
where $b(\cdot)$ denotes the so-called belief, $\mathcal{H}(b_{n})$
is the entropy of the belief of the $n$-th variable,
$\phi_{n}(x_{n})$ and $\psi_{m}(x_{C_{m}})$ are the
\textit{potential} function and the
\textit{compatibility} function, respectively, which are defined
depending on the application. Notice that if we use $C_{m}$ to denote
the $m$-th node's cluster ($m=1,2,\ldots,M$), then
$b_{C_{m}}(x_{C_{m}})$ denotes joint belief and
$\mathcal{I}_{C_{m}}(b_{C_{m}})$ represents joint mutual information
term \cite{Cover}. The optimization with respect to
$(\boldsymbol{b},\boldsymbol{\rho})$ starts with a fixed
$\boldsymbol{\rho}^{(k)}$, then solves this for the stationary points of
$\mathcal{F}(\boldsymbol{b},\boldsymbol{\rho}^{(k)})$ via TRW-BP. Next,
for a fixed belief vector $\boldsymbol{b}$, minimizing the function
$\mathcal{F}(\boldsymbol{b},\boldsymbol{\rho}^{(k)})$ with respect to
$\boldsymbol{\rho}^{(k)}$ results in an updated
$\boldsymbol{\rho}^{(k+1)}$. This algorithm keeps running
recursively until the belief converges. Observe that standard BP corresponds to the sub-optimal and generally invalid choice $\boldsymbol{\rho}=\mathbf{1}$.
The work reported in
\cite{Wainright2} is not directly applicable to
problems such as the decoding of LDPC codes, and only
derives message-passing rules for graphs with pairwise interactions.

In \cite{Wymeersch2}, \cite{Wymeersch}, the uniformly reweighted
(URW)-BP algorithm extends pairwise factorizations to higher order
interactions and reduces a series of globally optimized parameters
$\boldsymbol{\rho}\in(0,1]^{M}$ to a simple constant
$\rho_{u}\in(0,1]$. Additionally, the FAPs are generalized to edge
appearance probabilities (EAPs) so that the problem size is
significantly reduced. Another reweighting strategy is referred to
as variable FAP (VFAP)-BP \cite{LiuLam:12} that aims to select $\boldsymbol{\rho}$ on the basis of the cycle distribution of the
graph. However, neither URW-BP nor VFAP-BP optimizes the values of
$\boldsymbol{\rho}$ explicitly.

\begin{table}[!t]
\centering \caption{\label{tab:Algorithm-Flow-of-1}}LOW-BP for Decoding
LDPC Codes

\begin{small} %
\begin{tabular}{ll}
\hline & \tabularnewline

\textbf{{Offline: subgraphs formation} } & \tabularnewline
 & \tabularnewline
1: Given an expansion strategy (\textit{disjoint} or \textit{RA}) and $d_{\mathrm{max}}$,&
\tabularnewline apply PEG expansion to generate
$T\ge1$ subgraphs; & \tabularnewline
 & \tabularnewline
\textbf{{Offline: optimization of $\boldsymbol{\rho}_{t}$ for the
$t$-th subgraph}} & \tabularnewline
 & \tabularnewline
2: Initialize $\boldsymbol{\rho}_{t}^{(0)}$ to an allowable value;  & \tabularnewline
 & \tabularnewline
3: For each subgraph, calculate the beliefs $b(\boldsymbol{x}_t)$ and
& \tabularnewline the mutual information term
$\boldsymbol{I}_{t}=[I_{t,1},I_{t,2},\ldots,I_{t,L_{t}}]$ &
\tabularnewline by using reweighted message passing rule \eqref{mp1}--\eqref{mp3}; \\

 & \tabularnewline
4: With $b(\boldsymbol{x}_t)$ and $\boldsymbol{I}_{t}$ obtained from
step 3, update  & \tabularnewline $\boldsymbol{\rho}_{t}^{(r)}$ to
$\boldsymbol{\rho}_{t}^{(r+1)}$ using the conditional gradient method &
\tabularnewline (detailed in the Appendix);\\

 & \tabularnewline
5: Repeat steps 3--4 until $\boldsymbol{\rho}_{t}$ converges for
each subgraph; & \tabularnewline
 & \tabularnewline
\textbf{Offline: choice of $\boldsymbol{\rho}=[\rho_{1},\rho_{2},\ldots,\rho_{M}]$ for
decoding}  & \tabularnewline
 & \tabularnewline
6: For all $T$ subgraphs, collect $\boldsymbol{\rho}_{1},\ldots,
\boldsymbol{\rho}_{i}, \ldots, \boldsymbol{\rho}_{T}$.  &
\tabularnewline In case of disagreement on a value $\rho_{m}$ for the $m$-th & \tabularnewline check node,
choose   the value offering
the best performance; & \tabularnewline
 & \tabularnewline
\textbf{{Online: real-time decoding }} & \tabularnewline
 & \tabularnewline
7: Use reweighted message passing decoding \eqref{mp1}--\eqref{mp3} with & \tabularnewline  optimized $\boldsymbol{\rho}=[\rho_{1},\rho_{2},\ldots,\rho_{M}]$ during actual data transmission. \\

& \tabularnewline
\hline & \tabularnewline
\end{tabular}\end{small}
\end{table}

\section{Proposed LOW-BP Algorithm for Decoding LDPC Codes}

In this section, we describe LOW-BP, which explicitly optimizes the reweighting parameter vector
$\boldsymbol{\rho}=[\rho_{1},\rho_{2},\ldots,\rho_{M}]$. By
allowing optimization over smaller subgraphs, LOW-BP is able to
trade off complexity vs.~performance. LOW-BP comprises an offline
phase, during which, for a fixed SNR and a fixed code, the best
choice of $\boldsymbol{\rho}$ is determined, as outlined in Table
\ref{tab:Algorithm-Flow-of-1}. The online phase of LOW-BP occurs
during real-time decoding, when optimized $\boldsymbol{\rho}$ is
used in the reweighted message passing decoding algorithm.

\subsection{Offline Phase of LOW-BP}

In the offline phase, we transform the factor graph into a set of
$T\ge1$ subgraphs and then locally optimize the
reweighting parameter vector $\boldsymbol{\rho}_t$ for each subgraph,
where $t=1,2, \dots, T$. Note that when $T>1$ the dimension of
$\boldsymbol{\rho}_t$  depends on the size of the $t$-th subgraph. The optimization turns out to be significantly less complex
when more subgraphs are considered, hence there is a need for a flexible
method to decompose the original factor graph into subgraphs. We apply the
progressive-edge growth (PEG) technique \cite{Hu} to this end.

\subsubsection*{Construction of $T$ Subgraphs}

\begin{algorithm}
\caption{\label{cap:PEGalgo}
Progressive edge-growth (PEG) expansion for the $t$-th subgraph ($t=1,2, \ldots, T$)
}

\begin{algorithmic}[1]

\WHILE{ the complementary set $V_t$ is not empty}

\FOR{$j=0\to N-1$}

\FOR{$k=0\to w_{s_{j}}-1$}

\IF{ $k=0$}

\STATE add the first edge$(c_{i},s_{j})$ denoted by $e_{s_{j}}^{0}$
to $s_{j}$, such that $c_{i}$ has the lowest degree under the current
subgraph

\ELSE

\STATE keep expanding the subgraph from root $s_{j}$ and remove
$(c_{i},s_{j})$ from ${V}_t$ until the maximum level $d_{\mathrm{max}}$
is reached or $\bar{\mathcal{N}}_{s_{j}}^{d}\neq\emptyset$ but $\bar{\mathcal{N}}_{s_{j}}^{d+1}=\emptyset$

\ENDIF

\ENDFOR

\ENDFOR

\ENDWHILE

\end{algorithmic}
\end{algorithm}

PEG expands $\mathcal{G}(V,E)$ into $T$ subgraphs
$\mathcal{G}_{t}(V_t,E_t)$, where $V_t$ and $E_t$ are subsets of $V$ and
$E$, respectively. This method is straightforward to use if the
LDPC code was designed by PEG, or its variations \cite{uchoa,Xiao},
but is not limited to such designs. We consider a \textit{disjoint} strategy or a \textit{re-appearance
(RA)} strategy, to apply PEG.
The disjoint strategy
prohibits duplicates of check nodes in all subgraphs, while the RA
strategy allows check nodes to appear more than once over $T$
subgraphs. In general, the number of subgraphs $T$ depends on: (i) a
pre-set maximum expansion level $d_{\mathrm{max}}$, as a large
$d_{\mathrm{max}}$ results in a small $T$ but a high probability of the
existence of very short cycles within subgraphs; (ii) given
$d_{\mathrm{max}}$, whether all the nodes of $V$ are included in the
expanded subgraphs. Let us denote the degree of a variable node
$s_{j}$ by $w_{s_{j}}$, and define $\mathcal{N}_{s_{j}}^{d}$ as the
neighborhood of $s_{j}$ at current expansion level $d$, as well as
$\bar{\mathcal{N}}_{s_{j}}^{d}$ being the complement of
$\mathcal{N}_{s_{j}}^{d}$. To generate the $t$-th subgraph
$\mathcal{G}_{t}(V_t,E_t)$ based on $\mathcal{G}(V,E)$, the PEG
expansion is detailed in Algorithm \ref{cap:PEGalgo}. In the case of
the \textit{RA} strategy, ${V_t}$, the set of candidate nodes of
$\mathcal{G}_{t}(V_t,E_t)$, is always initialized as $V$, for each
of the $T$ expansions. On the other hand, ${V_t}$ is the complement
set of ${V_{t-1}}$ if the \textit{disjoint} strategy is applied, so
that the size of subgraph $\mathcal{G}_{t}(V_t,E_t)$ decreases as
$t$ increases. Furthermore, some of the subgraphs, such as
$\mathcal{G}_{T}(V_{T},E_T)$ or
$\mathcal{G}_{T-1}(V_{T-1},E_{T-1})$, may be a acyclic (i.e., a
tree), with corresponding reweighting factors
$\boldsymbol{\rho}_t=\boldsymbol{1}$, which complies with the
observations in \cite{Wainright2}, \cite{Wymeersch2}. Compared to
the greedy search algorithm in \cite{Hu,uchoa}, our PEG-based
expansion stops as soon as every member of $V_t$ has been visited.
The number of edges incident to $s_{j}$ might be less than
$w_{s_{j}}$ as some short cycles are excluded from the subgraphs so as to
guarantee that the local girth of each subgraph is always larger
than the global girth of the original graph.

\subsubsection*{Optimization of FAPs}

After obtaining $T$ subgraphs, we introduce $\boldsymbol{L}=[L_{1},L_{2},\ldots,L_{T}]$, where
$L_{t}$ is the number of check nodes (possibly with duplicates) in
the $t$-th subgraph. Note that $\sum_{t}L_{t}=M$ if
a disjoint expansion is used, while
$\sum_{t}L_{t}>M$ if  the RA expansion is employed.
Observe also that when $T=1$, then $L=M$. With the
 $t$-th subgraph, we optimize the associated  FAPs $\boldsymbol{\rho}_t=[\rho_{t,1},\rho_{t,2},\ldots,\rho_{t,L_t}]$  using the optimization method for TRW-BP \cite{Wainright2}, modified to  higher-order interactions,
with the corresponding message-passing rules from \cite{Wymeersch}.
The optimization problem is solved recursively as follows, starting
from the initial values $\boldsymbol{\rho}_{t}^{(0)}$ for each of
the $T$ subgraphs:
\begin{enumerate}
\item For all $T$ subgraphs in parallel, for fixed $\boldsymbol{\rho}_{t}^{(r)}$,
use message-passing to calculate the beliefs $b(\boldsymbol{x}_t)$ and
the mutual information term
$\boldsymbol{I}_{t}=[I_{t,1},I_{t,2},\ldots,I_{t,L_{t}}]$, provided
with $L_{t}\le M$ check nodes in the $t$-th subgraph.
\item For all $T$ subgraphs in parallel, given
$\{\boldsymbol{I}_{t}\}_{t=1}^{T}$, use the conditional gradient
method (see the Appendix) to update, for all $t$, $\boldsymbol{\rho}_{t}^{(r)}$
to $\boldsymbol{\rho}_{t}^{(r+1)}$, then go back to step 1).
\end{enumerate}
When applying the RA strategy, it is possible for a check node to
have a non-unique FAP. In that case, we choose the FAP that gives
the best BER performance through simulations. For clarity, the two
steps above are referred to as a recursion. The optimization of the
FAPs runs for as many recursions as are necessary until each
$\boldsymbol{\rho}_{t}$ converges, in order to acquire the optimal
reweighting vector.

\subsection{Online Phase of LOW-BP}

Once the optimized values of $\boldsymbol{\rho}=[\rho_{1},\rho_{2},\ldots,\rho_{M}]$ are found, actual
data transmission can commence. For completeness, we briefly review the message-passing
rules applied here, and refer the reader to \cite{Wymeersch} for more details. All messages are represented in log-lkelihood ratios (LLRs). For an AWGN channel with noise variance
$\sigma^{2}$, the message from the $n$-th variable node $s_n$ to the $m$-th check node $c_m$ is given by
\begin{equation} \label{mp1}
\centering {\Psi_{nm}=\lambda_{\mathrm{ch},n}+\sum_{m'\in \mathcal{N}(n)\backslash
m}\rho_{m'}\Lambda_{m'n}-(1-\rho_m)\Lambda_{mn}},
\end{equation}
where $\lambda_{\mathrm{ch},n}=\log (p(y_n|x_n=1)/p(y_n|x_n=0))=2 y_n/\sigma^{2}$, $m'\in \mathcal{N}(n)\backslash m$ is the neighboring set of check nodes of $s_n$ except $c_m$. The quantity $\Lambda_{mn}$
denotes messages sent from $c_m$ to $s_n$ in previous decoding iteration,
then for check nodes $c_m$ we
update $\Lambda_{mn}$ as:
\begin{equation} \label{mp2}
\centering {\Lambda_{mn}= f_{\boxplus}\big{(}\{\rho_{m}\Psi_{nm'}\}_{m'\in \mathcal{N}(n)\backslash
m}\big{)}-(1-\rho_m)\Psi_{nm}},
\end{equation}
where $f_{\boxplus}(\cdot)$ denotes the standard BP message passing
rule to compute an LLR message from check node $c_m$ to variable
node $s_n$. The function $f_{\boxplus}(\cdot)$ can be implemented by
using the well-known hyperbolic tangent expressions
\cite{LiuLam:12}, or the numerically more stable Jacobian logarithm
\cite{Wymeerschbook, robertson1995comparison}. Upon convergence, we
have the belief $\lambda_{\mathrm{belief},n}$ with respect to $x_n$
given by
\begin{equation} \label{mp3}
\centering
\lambda_{\mathrm{belief},n}=\lambda_{\mathrm{ch},n}+\sum_{m\in \mathcal{N}(n)}\rho_m\Lambda_{mn}.
\end{equation}
It should be noted that in \eqref{mp1}--\eqref{mp3}, the standard BP
algorithm corresponds to $\rho_m=1, \forall m$. The receiver
utilizes the above message-passing rules and does not update
$\boldsymbol{\rho}$ as long as the channel conditions are unchanged.

\section{Simulation Results\label{sec:Simulation-Results}}

\begin{figure}
\centering{}\global\long\def\epsfsize#1#2{1.0\columnwidth}
 \epsfbox{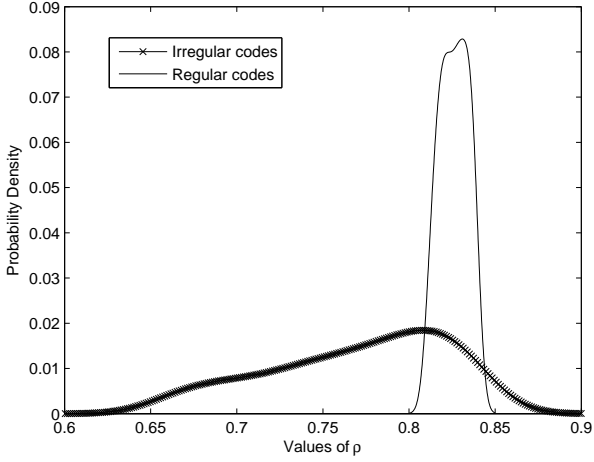} \vspace{-1.5em}
 \caption{\label{fig:Probability-densities-of}Histograms of the $\boldsymbol{\rho}$
values for regular codes and irregular codes at an  SNR of 2 dB. The $\boldsymbol{\rho}$ is derived by using LOW-BP optimization with disjoint selection and run until convergence. }
\end{figure}
In this section, we show the numerical results obtained from applying
the proposed LOW-BP algorithm to the decoding of regular and irregular LDPC
codes with short block lengths, over the additive white Gaussian noise
(AWGN) channel. The regular code has block length $N=500$ and rate
$R=1/2$, with constant column weight $w_{s}=4$ and row weight $w_{c}=6$.
The irregular code has the same block length and rate, but a variable
node degree distribution $\lambda(x)=0.21\times x^{5}+0.25\times x^{3}+0.25\times x^{2}+0.29\times x$
and a constant check node degree of 5.
For the sake of numerical stability
and data storage, all messages are represented as LLRs, and the Jacobian logarithm \cite{Wymeerschbook, robertson1995comparison} is used to compute the messages passed
from check nodes to variable nodes. In the offline phase of LOW-BP, $1000$ codewords known to the receiver are transmitted  so as to optimize $\boldsymbol{\rho}=[\rho_{1},\rho_{2},\ldots,\rho_{M}]$. We allowed up to
60 decoding iterations in the online phase.

Fig.~\ref{fig:Probability-densities-of} illustrates the distribution
of the reweighting parameters for regular codes and irregular codes,
at an SNR of 2 dB. It is clear that the optimized
$\boldsymbol{\rho}$ of irregular codes is widely distributed over
the range of $[0.6,0.9]$, while the $\boldsymbol{\rho}$-distribution
for regular codes is more concentrated within a smaller range
$[0.8,0.85]$. This observation is congruent with the findings in
\cite{Wainright2,Wymeersch2}, which state that for symmetric graphs,
the optimal reweighting parameters should be more or less uniform.

\begin{figure}
\centering{}\global\long\def\epsfsize#1#2{1.0\columnwidth}
 \epsfbox{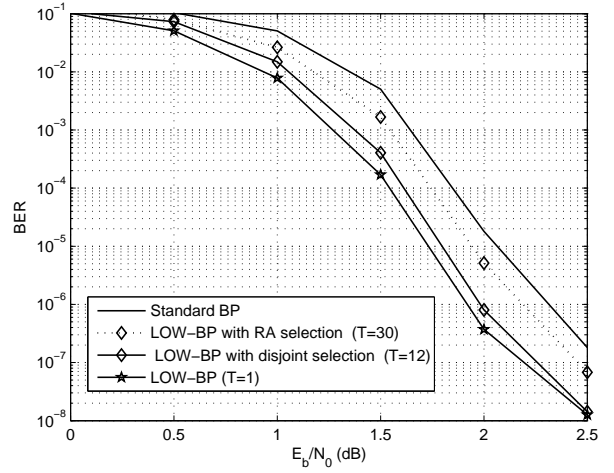} \vspace{-1.5em}
 \caption{\label{fig:treeregular}Comparison of decoding performance using the
proposed LOW-BP algorithm with various numbers of subgraphs $T$ for
regular codes.}
\end{figure}

\begin{figure}
\centering{}\global\long\def\epsfsize#1#2{1.0\columnwidth}
 \epsfbox{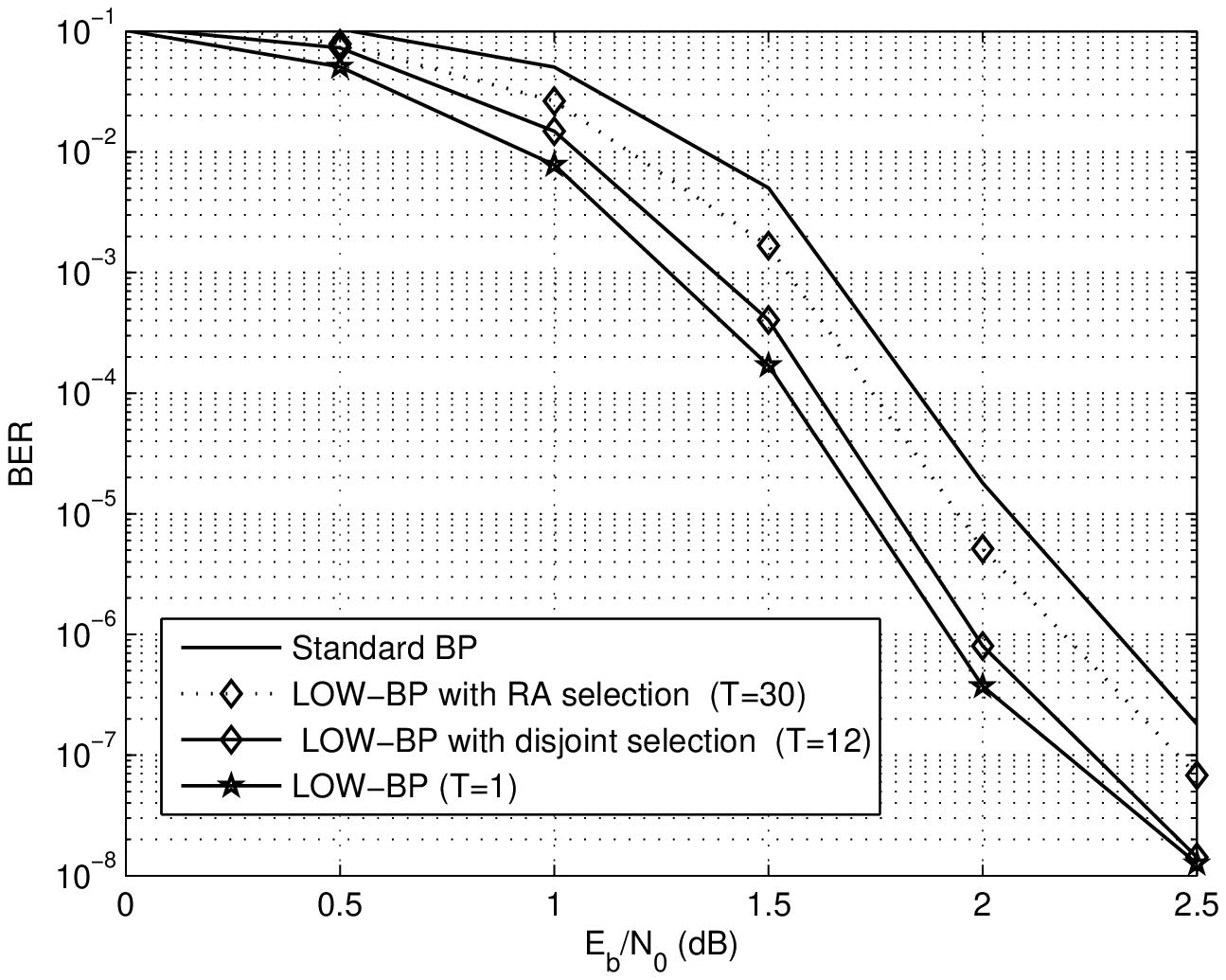} \vspace{-1.5em}
 \caption{\label{fig:treeirregular}Comparison of decoding performance using
the proposed LOW-BP algorithm with various numbers of subgraphs $T$
for irregular codes.}
\end{figure}

\begin{figure}
\centering{}\global\long\def\epsfsize#1#2{1.0\columnwidth}
 \epsfbox{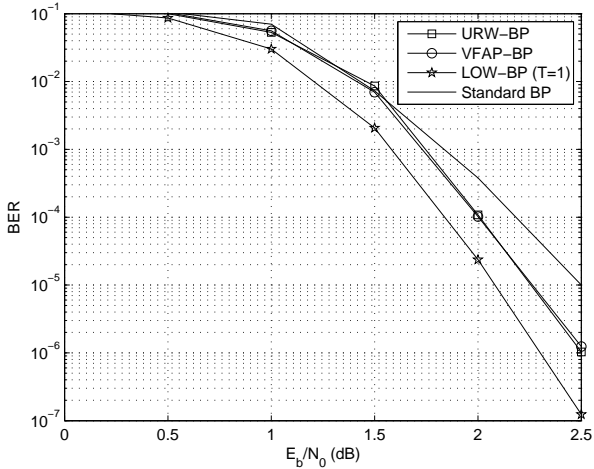} \vspace{-1.5em}
 \caption{\label{fig:regularber}Comparison of the performance of BP, URW-BP, VFAP-BP and the proposed LOW-BP for decoding regular codes. }
\end{figure}

\begin{figure}
\centering{}\global\long\def\epsfsize#1#2{1.0\columnwidth}
 \epsfbox{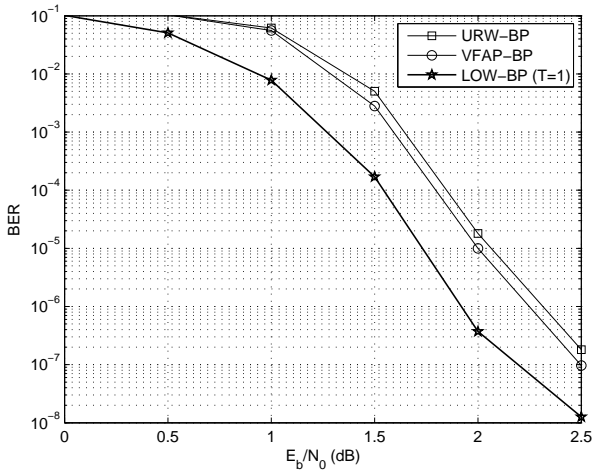} \vspace{-1.5em}
 \caption{\label{fig:irregularber}Comparison of the performance of BP (same results
as URW-BP), URW-BP, VFAP-BP and the proposed LOW-BP for decoding
irregular codes. }
\end{figure}

In Fig.~\ref{fig:treeregular} and Fig.~\ref{fig:treeirregular}, the
BER performance with different values of $T$, with RA and disjoint
selection, are compared to the standard BP algorithm, for regular
and irregular codes, respectively. We observe a performance gain of
up to $0.4$ dB over the standard BP algorithm by using the proposed
LOW-BP method. For regular code, all check nodes are visited
once in $T=9$ subgraphs with the disjoint selection, while with RA
selection $T=25$ subgraphs are generated (maximum 60 recursions),
where some check nodes are revisited. For irregular code, the
disjoint selection generates $T=12$, meanwhile the RA selection gives
$T=30$ (maximum 100 recursions). When using disjoint selection,
$\boldsymbol{\rho}$ converges to a set of stable values after a
number of recursions that varies from one subgraph to another.
Notice that in both figures, $T=1$ is a special case that
corresponds to Wainwright et al.'s optimal solution from
\cite{Wainright2}. For $T=1$, to improve convergence in the offline
phase, we initialized $\boldsymbol{\rho}$ from URW-BP
\cite{Wymeersch2} for regular code and from VFAP-BP
\cite{LiuLam:12} for irregular code. Normally, around $800$
recursions we required to converge for regular code and $2700$
recursions for irregular code.

A comparison with existing reweighted methods is shown in
Fig.~\ref{fig:regularber} and Fig.~\ref{fig:irregularber}, for regular and irregular codes, respectively. The algorithms considered are URW-BP from \cite{Wymeersch2}, VFAP-BP from
\cite{LiuLam:12}, and the proposed LOW-BP algorithm (with $T=1$).
For regular code, we observe that URW-BP and VFAP-BP outperform
standard BP. LOW-BP is able to provide further improvements. For irregular code, the optimal constant value of the FAP in URW-BP is
$\rho=1$, so that BP and URW-BP coincide. VFAP-BP provides a small
performance gain, while LOW-BP again outperforms BP by up to $0.4$
dB. We clearly see that explicit optimization of $\boldsymbol{\rho}$
leads to non-trivial performance gains.

\section{Conclusion}

We have proposed a locally-optimized reweighting belief propagation
(LOW-BP) algorithm for decoding finite-length LDPC codes. The
proposed algorithm has been compared to previously reported
reweighted belief propagation algorithms and has demonstrated superior
performance for the scenarios considered. LOW-BP comprises an
offline and an online stage. The online stage relies on standard
tree-reweighted belief propagation, while the offline stage involves
an optimization problem over subgraphs of the original factor graph.
By increasing the number of subgraphs, the offline stage converges
faster and exhibits less complexity. Reducing the number of
subgraphs will lead to improved BER performance, albeit at an additional
delay and complexity cost during the offline stage. LOW-BP is
especially well suited to decoding of short to moderate LDPC codes and is a promising choice for applications that require a reduced
number of decoding iterations. Future avenues of research include
fast adaptation of the offline stage to time-varying channel
conditions.

\section*{Appendix: Details of Conditional Gradient}

Our goal is to minimize \eqref{objective} with respect to the column vector $\boldsymbol {\rho}_t$, for  a specific subgraph $\mathcal{G}_{t}(V_t,E_t)$. Dropping terms that do not depend on $\boldsymbol {\rho}_t$, we find the following
optimization problem with $\boldsymbol{I}_{t}=[I_{t,1}~I_{t,2}\ldots I_{t,L_t}]^{\dagger}$, where $(\cdot)^{\dagger}$ denotes the transpose:
\begin{align*}
\mathrm{minimize} & \,\,\,\,-\boldsymbol {\rho}_t^{\dagger}\boldsymbol{I}_{t}\\
\mathrm{s.t.} & \,\,\,\,\boldsymbol{\rho}_{t} \in \mathbb{T}\big{(}\mathcal{G}_{t}(V_t,E_t)\big{)},
\end{align*}
where $\mathbb{T}\big{(}\mathcal{G}_{t}(V_t,E_t)\big{)}$ is the set of all valid FAPs over
the subgraph $\mathcal{G}_{t}(V_t,E_t)$ and $I_{t,l}$ is a mutual information term, which depends on $\boldsymbol {\rho}^{(r)}_t$, the previous value of $\boldsymbol {\rho}_t$.  We will denote the objective function by $f(\boldsymbol {\rho}_t)=-\boldsymbol {\rho}_t^{\dagger}\boldsymbol{I}_{t}$ and use the conditional gradient method to update $\boldsymbol {\rho}_t$, similar to \cite{Wainright2}. In the conditional gradient method, we first linearize the objective around the current value $\boldsymbol {\rho}^{(r)}_t$:
\begin{equation}\label{flin}
f_{\mathrm{lin}}(\boldsymbol {\rho}_t)=f(\boldsymbol {\rho}^{(r)}_t) + \nabla_{\boldsymbol {\rho}_t}^{\dagger}f(\boldsymbol {\rho}^{(r)}_t) (\boldsymbol {\rho}_t-\boldsymbol {\rho}^{(r)}_t),
\end{equation}
in which $\nabla_{\boldsymbol {\rho}_t}f(\boldsymbol {\rho}^{(r)}_t)=-\boldsymbol{I}_{t}$. Secondly, we minimize $f_{\mathrm{lin}}(\boldsymbol {\rho}_t)$ with respect to $\boldsymbol {\rho}_t$, denoting the minimizer by $\boldsymbol{\rho}_t^{\ast}$ and $z^{(r+1)}_t=\max(f_{\mathrm{lin}}(\boldsymbol{\rho}_t^{\ast}),z^{(r)}_t)$, where $z^{0}_t=-\infty$.  Finally, $\boldsymbol {\rho}^{(r)}_t$ is updated to $\boldsymbol {\rho}^{(r+1)}_t$ as
\begin{equation}\label{11}
\centering{\boldsymbol{\rho}_t^{(r+1)}=\boldsymbol{\rho}_t^{(r)}+\alpha(\boldsymbol{\rho}_t^{\ast}-\boldsymbol{\rho}_t^{(r)})},
\end{equation} in which $\alpha$ is chosen as
\begin{equation} \label{10}
\arg \min_{\alpha \in [0,1]}f(\boldsymbol{\rho}_t^{(r)}+\alpha(\boldsymbol{\rho}_t^{\ast}-\boldsymbol{\rho}_t^{(r)})).
\end{equation}
At every iteration, $f(\boldsymbol{\rho}_t^{(r)})$ is an upper bound on the optimized objective, while $z_t^{(r+1)}$ is a lower bound. 

\section*{Acknowledgment}

This work was supported, in part, by the European Research Council,
under grant No. 258418.

\end{document}